\documentclass[twoside,fleqn]{article}
\usepackage{espcrc2}
\usepackage{times,mathptm}
\usepackage[dvips]{graphicx}
\newcommand{\oh}{\frac{1}{2}}

\newcommand{\beq}{\begin{equation}}
\newcommand{\eeq}{\end{equation}}
\newcommand{\bea}{\begin{eqnarray}}
\newcommand{\eea}{\end{eqnarray}}
\newcommand{\be}{\begin{equation}}
\newcommand{\ee}{\end{equation}}
\newcommand{\A}{{\cal A}}
\title{Center vortices at $N>4$ colors%
\thanks{Presented by J.\ Greensite.  Supported in part by 
the U.S.\ Department of Energy under Grant No.\ DE-FG03-93ER40711.
Our collaborative effort was also supported by NATO Collaborative Linkage 
Grant No.\ PST.CLG.976987.}}
\author{%
Jeff Greensite\address{Physics and Astronomy Dept., San Francisco State Univ.,
San Francisco, CA~94117, USA}%
$^,$\address{The Niels Bohr Institute,
DK--2100 Copenhagen \O, Denmark}
and 
\v{S}tefan Olej\-n{\'\i}k\address{Institute of Physics, Slovak Academy 
of Sciences, SK--842 28 Bratislava, Slovakia}}
\begin{document}
%
%
\begin{abstract}
	We discuss two issues related to the physics of center vortices in pure
SU($N$) lattice gauge theory at large $N$: 
{\bf 1.}\ Center vortices are stable classical solutions
of the Wilson action, as well as of a wide class of improved lattice actions,
for any $N>4$.
{\bf 2.}\ The natural scaling of $k$-string tensions at large $N$, 
in the vortex picture of confinement, is 
$\sigma(k)=k\sigma(1)$. This is the common large $N$ limit of Casimir and
Sine Law scaling. The crucial feature for explaining this
behavior is the existence of center monopoles.
\end{abstract}
\maketitle
%
%
\section{CENTER VORTICES AS CLASSICALLY\\
STABLE OBJECTS FOR $N>4$ COLORS}\label{minima}
	Lattice simulations provide abundant evidence for the
important role of center vortices in the mechanism of color 
confinement~\cite{DelDebbio:1998uu}. 
However, it is generally believed that unlike instantons, 
which are known to be stable local minima of the pure  SU($N$) 
gauge theory action, center vortices exist as stable classical 
solutions only in the case of gauge fields coupled to a set of adjoint 
Higgs fields, when the symmetry is broken from SU($N$) to $Z_N$.

	Surprisingly, there is no need to break the SU($N$) symmetry
to obtain stable vortex solutions on the lattice. The following
remarkable result, by now almost forgotten, was obtained by Bachas 
and Dashen in 1982~\cite{Bachas:1982aa}: \textit{Thin center vortices are stable local minima of the 
Wilson lattice action for any SU($N$) pure gauge theory with $N>4$.}
We will briefly recall the proof of this statement, report on its extension
to a wide class of two-parameter improved actions, and comment on its 
importance in the context of the renormalization group.

\subsection{Wilson action}
The proof for the Wilson action is rather trivial:
Any thin center vortex in SU($N$) lattice gauge theory 
is gauge equivalent to
\begin{equation}\label{thin}
U_\mu(x)\ =\ Z_\mu(x) I_N
\end{equation}
with
\[
\nonumber
Z_\mu(x)=\exp\left(\frac{2\pi i n_\mu(x)}{N}\right),
\quad
n_\mu(x)=1, 2, \dots, N-1.
\]
Now write a small deformation of this configuration as
\begin{equation}\label{deformation}
U_\mu(x)=Z_\mu(x) V_\mu(x), 
	\qquad V_\mu(x)=e^{i A_\mu(x)}
\end{equation}
with
\begin{equation}
	A_\mu(x)=\sum_a A_\mu^a(x)L_a,\qquad \vert A_\mu^a(x)\vert \ll 1.
\end{equation}
Substituting (\ref{deformation}) into the Wilson action
\begin{equation}
S=\frac{\beta}{2N}\sum_P 
	\left(2N-\mbox{Tr}[U_P]-\mbox{Tr}[U^\dagger_P]\right)
\end{equation}
we obtain
\begin{equation}
S=\frac{\beta}{2N}\sum_P 
	\left(2N-Z_P\mbox{Tr}[V_P]-Z^\ast_P\mbox{Tr}[V^\dagger_P]\right)
\end{equation}
Writing the product of $V$-link variables around a plaquette in terms of
a field strength
\begin{equation}
V_P=e^{i F_P}=I_N+i F_P -\oh F_P^2 +{\cal{O}}(F_P^3),
\end{equation}
and substituting for $Z_P$
\begin{equation}
Z_P=\exp\left(\frac{2\pi i n_P}{N}\right)
\end{equation}
one gets
\begin{eqnarray}
\nonumber
S=\frac{\beta}{2N}\sum_P
	\left[2N\left(1-\cos\left(\frac{2\pi  n_P}{N}\right)\right)\right.
\\
\left.+\cos\left(\frac{2\pi  n_P}{N}\right)
	\mbox{Tr}[F_P^2]
	+{\cal{O}}(F_P^3)\right].
\end{eqnarray}
The action has a \textit{local minimum} at 
$\mbox{Tr}[F_P^2]=0$
\textbf{\textit{providing}} that at each plaquette
	with {$n_P>0$}
\begin{eqnarray}
\cos\left(\frac{2\pi n_P}{N}\right)>0,&&\\
\mbox{i.e.\ }\quad\frac{n_P}{N}<\frac{1}{4}
	\quad&\mbox{\ or\ }&\quad
	\frac{N-n_P}{N}<\frac{1}{4}.
\label{condition}
\end{eqnarray}
This vortex stability condition cannot be satisfied for $N\le4$, however, 
beginning with $n_P=1, 4$ at $N=5$, vortex stability is obtained from the
Wilson action already at the classical level.

\subsection{Improved actions}
  The above result can be extended to more complicated lattice actions,
and therein lies its physical relevance.
Thin vortices, stable or not, are suppressed at weak couplings by a
factor of order $\exp(-\mbox{Vortex Area}/g^2)$ and do not percolate. 
The configurations of physical interest are center vortices
having some finite thickness in physical units. 
To investigate the stable classical 
configurations with a certain length scale $d$, starting from a lattice 
action at 
spacing $a$ and including quantum fluctuations up to the scale $d$, we can 
follow the RG approach and apply successively blocking transformations
\begin{equation}
e^{-S'[{\cal{U}}]}=\int DU\;\delta[{\cal{U}}-F(U)]e^{-S[U]},
\end{equation}
where $\cal{U}$ are links on the blocked lattice, and $F(U)$ is the blocking
function.

	In this kind of approach, one usually assumes that only a few contours
(plaquettes, 6- and 8-link loops, etc.) are important in the effective 
action. We have studied, in particular, the class of two-parameter improved
lattice actions which consist of plaquette and $1\times 2$ rectangle terms:
\begin{eqnarray}
\nonumber
S_I&=&c_0{\sum_{P}}(N-\mbox{ReTr}[U(P)])\\
&+&c_1{\sum_{1\times 2\ R}}(N-\mbox{ReTr}[U(R)]).\label{improved}
\end{eqnarray}
This simple extension beyond the Wilson action 
includes many lattice actions discussed in the literature
(tadpole-improved, Iwasaki, and DBW2 actions, two-parameter approximations
to the Symanzik action), and has been applied to MCRG studies of
the renormalization trajectory \cite{deForcrand:2000aa}. 

	The thin vortex configuration, eq.\ (\ref{thin}), is easily
shown to be a stable minimum
of the action $S_I$ providing both $c_0$ and $c_1$ are positive; the proof
is as simple as in the case
of the Wilson action. But for most improved actions of this type one has
 $c_0>0$ and $c_1<0$, and the proof of stability is a little more
involved.  We will only report here our basic result, which will be
derived in full in a subsequent publication \cite{Greensite:2002zz}:  
Thin center vortex configurations are stable local minima 
of the two-parameter action if, first,
the trivial vacuum $U_\mu(x)=I_N$ is the global minimum of the action,
which is satisfied iff
\begin{equation}
c_0+8 c_1>0 .
\end{equation}
The second condition for vortex stability is
the same as for the Wilson action, namely, 
that eq.\ (\ref{condition}) satisfied.

  So it seems that the result first obtained by Bachas and 
Dashen~\cite{Bachas:1982aa} is 
quite robust: Center vortices at $N>4$ are stable minima
of lattice actions in a large region of coupling constant space
associated with improved actions. Assuming they remain as local
minima all along the renormalization trajectory, their effects must 
become apparent at some scale. The reason is that the entropy
factor increases with vortex surface area as $\exp[+\mbox{const}\cdot
(\mbox{Vortex Area})]$, while the Boltzmann suppression factor
goes like $\exp[-(\mbox{Vortex Area})/\kappa^2(d)]$. As $d$ increases,
so does $\kappa^2(d)$. Eventually entropy wins over action, and
vortices at that scale will percolate through the lattice.

    At large distance scales, plaquette terms 
in the adjoint representation should
also appear and become important in the lattice effective action; such
terms can stabilize center vortices even in the $N\le 4$ case. 
This mechanism has been demonstrated explicitly in the context of 
strong-coupling lattice gauge theory in ref.~\cite{Faber:2000aa}.
%
%
\section{$k$-STRING TENSIONS AT LARGE $N$}\label{k-strings}
	For SU($N$) gauge theories with $N>3$ there are a number
of color representations in which color charge cannot be screened by gluons.
These unscreenable representations correspond to the lowest dimensional
SU($N$) representation with $N$-ality $k$. There are two predictions for
the $k$ dependence of string tensions between such color charges: 
\textbf{\textit{Casimir scaling}}
\begin{equation}
\sigma(k)=\frac{k(N-k)}{N-1}\sigma(1),
\end{equation}
based on dimensional-reduction arguments, and
\textbf{\textit{the Sine Law}}
\begin{equation}
\sigma(k)=\frac{\sin(\pi k/N)}{\sin(\pi/N)}\sigma(1),
\end{equation}
motivated by MQCD.

	In the large $N$ limit, Casimir scaling becomes exact due to the
factorization property. In this limit, the difference between the above
formulas disappears. Both predict, for $k$-string tensions with $k\ll N$,
\beq
\sigma(k)=k\sigma(1)
\eeq
which we refer to as \textbf{\textit{``$k$-scaling''}}. 
Is this property a feature of the
center vortex confinement mechanism?

	The affirmative answer to this question is based on the following
chain of arguments:

	1.\ The distribution of center vortices in an SU($N$) gauge theory is
very likely controlled by an effective $Z_N$ gauge theory, namely, the
P-vortex effective action.  Calculation of
center-projected observables in an SU($N$) theory fixed to an adjoint 
(e.g.\ maximal center) gauge is equivalent to calculation of 
observables in the effective $Z_N$ gauge
theory. These calculations have been carried out extensively in
the $N=2$ case \cite{DelDebbio:1998uu} (there are
some results for SU(3) as well).

	2.\ The effective $Z_N$ action is certainly non-local at the
lattice scale, but should be local at the color-screening scale. Its
excitations, at this scale, are thin center vortices and center
monopoles (for $N\ge 3$). For large $N$, the $Z_N$ gauge group
approximates a U(1) group, and center monopoles go over to the abelian
monopoles of compact QED. A Wilson loop of $N$-ality $k$ then becomes
a Wilson loop of $k$ units of abelian charge.

	3.\ Finally, the charge dependence of string tensions in the
U(1) theory can be calculated \`a la
Po\-lya\-kov~\cite{Polyakov:1977aa}, using the monopole Coulomb gas
representation. This was done, in $D=3$ dimensions, in
Ref.~\cite{Ambjorn:1998xx}; the result is precisely $k$-scaling.

     Invoking center dominance, $k$-scaling in
the effective $Z_N$ theory carries over to $k$-scaling of $k$-string
tensions in the full SU($N$) theory.  This is how the $k$-scaling property
at large $N$ is obtained in the center vortex confinement picture.
  
   We conclude with a comment about vortex densities at large $N$.
If we assume that center flux is essentially uncorrelated among
regions of area $\A>\A_{min}$ in a plane, then we can subdivide the
plane into square regions of area $\A_{min}$ and ask for the number of
these regions, per unit area, which are pierced by center flux of
magnitude $2\pi l/N$.  This number per unit area defines a vortex density
$\rho(l,N)$.  Given $k$-scaling, it is then
straightforward to calculate $\rho(l,N)$, and we find that it falls
like $1/N$ at large $N$.  This is in contrast to a recent result in
ref.\ \cite{D3}, which argues that a vortex density consistent with
$k$-scaling would have to \emph{increase} linearly with $N$.  We
believe that the discrepancy can be traced to a dilute gas
approximation used in ref.\ \cite{D3}, which appears to be
inconsistent with the existence of a lower bound $\A_{min}$ at large
$N$ \cite{Greensite:2002zz}.

\section*{Acknowledgment}
{J.G.\ thanks Herbert Neuberger for bringing 
Ref.~\cite{Bachas:1982aa} to his attention.}

%
%

\end{document}